\documentclass[prc,twocolumn,letterpaper,amsmath,amssymb,floatfix,nofootinbib]{revtex4}
\usepackage{graphicx,bm,url}
\pdfoutput=1
\newcommand{\mybar}[1]{\bar{#1\rule{0pt}{7.5pt}}}

\begin{document}

\title{Corrections to the neutrinoless double-beta-decay
operator in the shell model}

\author{Jonathan Engel} 

\affiliation{Dept.\ of Physics and Astronomy,  University of North Carolina, Chapel
Hill, NC, 27516-3255, USA}

\author{Gaute Hagen} 
\affiliation{Physics Division, Oak Ridge National Laboratory, P.O. Box 2008, Oak Ridge, TN 37831, USA. }

\begin{abstract}

We use diagrammatic perturbation theory to construct an effective shell-model
operator for the neutrinoless double-beta decay of $^{82}$Se.  The starting
point is the same Bonn-C nucleon-nucleon interaction that is used to generate
the Hamiltonian for recent shell-model calculations of double-beta
decay.  After first summing high-energy ladder diagrams
that account for short-range correlations and then adding diagrams of low order
in the $G$ matrix to account for longer-range correlations, we fold the
two-body matrix elements of the resulting effective operator with transition
densities from the recent shell-model calculation to obtain the overall
nuclear matrix element that governs the decay.  Although the high-energy ladder
diagrams suppress this matrix element at very short distances as expected, they
enhance it at distances between one and two fermis, so that their overall
effect is small.  The corrections due to longer-range physics are large, but
cancel one another so that the fully corrected matrix element is comparable to
that produced by the bare operator.  This cancellation between large and
physically distinct low-order terms indicates the importance of a reliable
nonperturbative calculation.

\end{abstract}


\maketitle
\section{Introduction} 

Neutrinoless double-beta decay is a very slow lepton-number-violating nuclear
process that occurs if neutrinos are their own antiparticles.  An initial
nucleus $(Z,A)$, with proton number $Z$ and total nucleon number $A$, decays to
$(Z + 2,A)$, emitting two electrons in the process \cite{avi08}.

The neutrino masses and mixing matrix figure prominently in the decay. The rate,
assuming that the process is mediated by the exchange of a light virtual
neutrino, is
\begin{equation}
\label{eq:finalrate}
[T^{0\nu}_{1/2}]^{-1}=G_{0\nu}(Q,Z)
  \left|M_{0\nu}\right|^2 \langle m_{\beta\beta} \rangle^2~,
\end{equation}
where $Q$ is the energy difference between the initial and final nuclei, $Z$ is
the charge of the initial nucleus, $G_{0\nu}(Q,Z)$ is a tabulated phase-space
measure, $M_{0\nu}$ is the nuclear matrix element to which we turn shortly, and
$m_{\beta\beta}$ is a linear combination of neutrino masses:
\begin{equation} \label{eq:effmass} m_{\beta\beta} \equiv |\sum_{k=1}^3 m_k
	U_{ek}^2|~.  \end{equation}
In this last equation, $m_k$ is the mass of the $k^\mathrm{th}$ neutrino (these
mass eigenstates are linear combinations of the electron-, mu- and
tau-neutrinos) and $U_{ek}$ is the element of the unitary mixing matrix that
connects that neutrino to the electron neutrino.   The quantity
$m_{\beta\beta}$ is what experimenters want to extract from the measured decay
rate.  They cannot do so, however, without knowing the matrix element
$M_{0\nu}$, which must be calculated in some nuclear model.

Most calculations of $M_{0\nu}$ are done either in the neutron-proton
Quasiparticle Random Phase Approximation (QRPA) or in the shell model.  The two
methods have complementary virtues.  The QRPA includes many single-particle
orbitals outside a relatively small ``inert'' core, but limits itself to a
particular kind of correlation.  The shell model includes arbitrary complicated
correlations, but only among a few single-particle orbitals outside a larger
inert core.  The current predictions of the two models, after a recent shaking
out period, show the QRPA matrix element exceeding that of the shell model by
factors of up to about two in the lighter isotopes such as $^{76}$Ge and
$^{82}$Se, and somewhat less in the heavier isotopes \cite{men08,sim08}.  

Which kind of calculation is closer to the truth?  Are there important effects
that escapes both models?  To find out, one has to correct one or both to
account for omitted physics. Although it is possible to add missing
correlations to the QRPA, it is not easy to do so systematically because
several different uncontrolled approximations --- BCS pairing, a
phenomenological interaction, the quasiboson approximation, etc.\ --- are part
of the method.  By contrast, because the shell model includes all correlations
within a well-defined subspace of the full Hilbert space (the space generated
by valence particles occupying a few single-particle states), there is a
systematic procedure for adding the effects of states outside that space
\cite{ell77,hjo95}.

While the procedure as usually implemented is perturbative in a renormalized
residual nuclear interaction (the $G$ matrix) and not always reliable for that
reason, it often works well enough, particularly if followed by some modest
adjustment to data.  Practitioners have long used such an approach to obtain
good effective interactions \cite{cau05}, but have never applied the same
techniques to obtain an effective double-beta-decay operator.  Instead, they
typically modify the bare operator phenomenologically, \emph{e.g.} through the
reduction of the axial-vector coupling constant $g_A$ (suggested by studies of
single-beta decay) or the use of a prescription \cite{mil76} to treat the
short-range nucleon-nucleon repulsion that is not present in shell-model wave
functions\footnote{The use of the Unitary Correlation Operator Method,
\emph{e.g}, in Ref.\ \cite{kor07,kor07a}, is more than a prescription, but the
method has not yet been consistently applied to both the decay operator and the
interaction.}.  Not surprisingly, it is difficult to assess the reliability of
such approximations.

In this paper, therefore, we apply the same techniques used to construct
effective shell-model interactions to the decay operator itself.  Section II
below contains a brief description of the matrix element we try to calculate,
and Section III a description of our procedure for renormalizing the
corresponding operator. In Section IV we present the results of our
calculation, which combines the renormalized operator with shell-model
transition densities from the authors of Ref.\ \cite{cau08} for the decay of
$^{82}$Se to $^{82}$Kr. (Densities for other decays, \emph{e.g.} of $^{76}$Ge,
are not currently available.)  Although we cannot be fully confident in our
perturbative result, our initial steps --- accounting for short-range
correlations through the generation of a $G$ matrix and an analogous corrected
decay operator --- are nonperturbative and trustworthy.  And even if the
low-order perturbation theory we employ subsequently is not accurate, it should
tell use whether we might expect significant renormalization in a fully
nonperturbative treatment. 

\section{Form of bare $M_{0\nu}$}

A precise expression for the matrix element is complicated, but with a few
approximations that induce an error of with an error of less than 30\%
\cite{men08,sim08}, we can write $M_{0\nu}$ as
\begin{equation}
\label{eq:approxme}
M_{0\nu} \approx M^{GT}_{0\nu}-\frac{g_V^2}{g_A^2}
M^{F}_{0\nu}
\end{equation}
with $g_V$ and $g_A$ the vector and axial-vector coupling constants, and
\begin{eqnarray}
   M^{F}_{0\nu}&=& \langle f |\sum_{a,b} H(r_{ab},\mybar{E}) \tau^+_a \tau^+_b
   |i\rangle\,, \label{eq:f}
\\
   M^{GT}_{0\nu}&=& \langle f |\sum_{a,b} H(r_{ab},\mybar{E}) \vec{\sigma}_a
\cdot \vec{\sigma}_b \tau^+_a \tau^+_b |i\rangle\,.\label{eq:gt}
\end{eqnarray}
Here $|i\rangle$ and $|f\rangle$ are the initial and final nuclear
ground states, $a,b$ label nucleons, $\mybar{E}$ is an average excitation
energy, and $H$ is a ``neutrino potential'', given by
\begin{equation}
   H(r,\mybar{E}) = \frac{2R}{\pi r} \int_0^{\infty} dq \frac{\sin{qr}}
   {\rule{0pt}{10pt}q + \mybar{E}-(E_i+E_f)/2}\,. \label{eq:poten}
\end{equation}
The quantity $R$ is the nuclear radius, inserted to make the matrix
element dimensionless.  Since our work is exploratory, we use the relatively
simple forms in Eqs.\ (\ref{eq:f}) and (\ref{eq:gt}) in most of what follows,
though we also discuss corrections due to nucleon form factors and forbidden
terms in the weak nuclear current.  

\section{Constructing an Effective Shell-Model Decay Operator}

\subsection{Formalism and diagrams for two-body operators}

Diagrammatic effective-operator theory has a long history in nuclear physics.
References \cite{bra67} and \cite{ell77} are early reviews and Ref.\
\cite{hjo95} is a more recent one.  The theory is based on the division of the
many-body Hilbert space into the shell-model space $P$ of particles occupying
several degenerate or quasidegenerate orbitals (usually eigenstates a
harmonic-oscillator potential $U$), and the rest of the Hilbert space $Q$.  One
begins by defining operators with the same names that project onto these spaces:
\begin{equation}
\label{eq:pq}
P=\sum_{i \ \epsilon P}|i \rangle\langle i | \,, \qquad \qquad
Q=\sum_{\mathrm{other} \ i} |i\rangle\langle i| \,,
\end{equation}
with
\begin{equation}
\label{eq:orthog}
P^2 = P\,, \qquad Q^2=Q\,, \qquad PQ = QP = 0\,.
\end{equation}
Next one defines an effective Hamiltonian that when acting on the $P$-space
projection of an eigenstate $|\Psi_a\rangle$ gives back that projection with
the correct eigenvalue:
\begin{equation}
\label{eq:heff}
 H_\mathrm{eff}(E_a) P |\Psi_a\rangle=   E_a P|\Psi_a\rangle \,.
\end{equation}
Similarly, for any ``bare'' operator $\mathcal{M}$ (where ``bare'' means
``acting in the full model space''), we can define an effective operator
$\mathcal{M}_\mathrm{eff}$ that acts only in the $P$ space, with matrix elements related to those of the bare
operator by: 
\begin{equation} 
\label{eq:meff}
 \frac{\langle \Psi_a | P \mathcal{M}_\mathrm{eff} P
|\Psi_b\rangle}{\sqrt{\langle \Psi_a|P|\Psi_a\rangle \langle
\Psi_b|P|\Psi_b\rangle}}= \langle \Psi_a |\mathcal{M}|\Psi_b\rangle
\end{equation}

It is straightforward to show that the effective Hamiltonian can be represented
in energy-dependent form as a solution to the Bloch-Horowitz equation
\cite{blo58}
\begin{equation}
\label{eq:bh}
H_\mathrm{eff}(E) = P H P + P H Q \frac{1}{E-QH}Q H P \,,
\end{equation}
with the full wave function a solution to the associated equation
\begin{equation}
\label{eq:bhwf}
|\Psi(E)\rangle = \mathcal{Z} \left( 1+\frac{1}{E-QH} QH \right)P|\Psi(E)
\rangle \,,
\end{equation}
Here $E$ is the energy of the eigenstate and $\mathcal{Z}$ is a normalization
factor.  One can remove the explicit dependence on energy by treating the
residual interaction $V-U$, which couples the $P$ and $Q$ spaces, as a
perturbation.  When Eqs.\ (\ref{eq:bh}) and (\ref{eq:bhwf}) are solved order by
order, the result is a series of valence-linked Goldstone diagrams for the
matrix elements matrix elements of $H_\mathrm{eff}$ and
$\mathcal{M}_\mathrm{eff}$ \cite{bra67}. 

The diagrams are very much like those for the binding energy, but have open
lines at each end to represent the valence single-particle states on the right
and left sides of the effective operators.  The use of this representation in a
linked-cluster expansion forces the introduction of ``folded'' diagrams, in
which intermediate states have zero excitation energy.  The mostly low-order
diagrams we consider here, however, will not have folds.

Some recent work on effective interactions \cite{cor08} has used
$V_\mathrm{low~k}$ as a starting point, with high-energy states effectively
integrated out at the beginning.  But because we want to calculate the
contributions of such states to the effective decay operator, we need a method
that treats them explicitly.  Our starting point, therefore, is the same as in
the traditional treatment of nuclear matter: we define a nonperturbative $G$
matrix as the sum of the two-particle ladder diagrams displayed in Fig.\
\ref{fig:fig1} below.  The ruled lines in the figure indicate high-energy
states, lying well above the shell-model single-particle space (we will vary
the exact amount by which they are above).  The $G$ matrix is thus defined not
only between two-body states in the valence space but also between states in a
larger model space that contains several higher shells.  (See Ref.\
\cite{hjo95} for details on this ``double partitioning'' of the Hilbert space.) 

\begin{figure}[htb]
\includegraphics{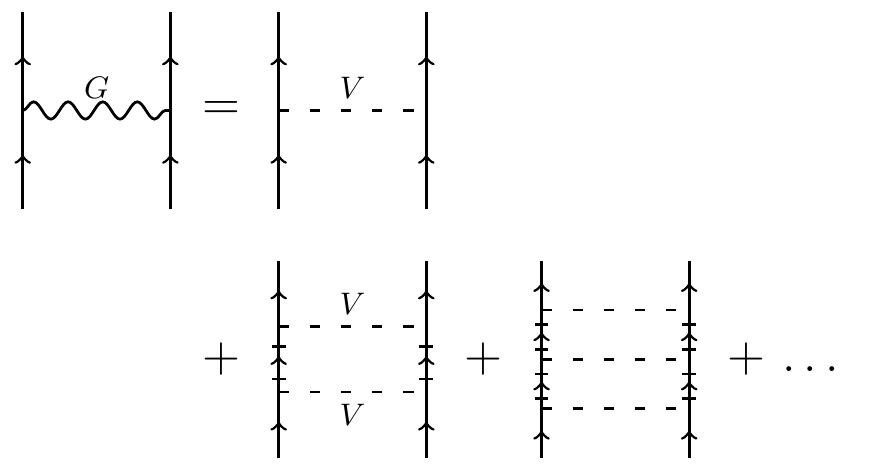}
\caption{\label{fig:fig1}Diagrammatic construction of a $G$ matrix, the first
step in an effective interaction.  Ruled lines correspond to high-energy
particle states in the doubly partitioned space.}
\end{figure}

Although there are an infinite number of ladder diagrams, the sum can be
carried out indirectly, \emph{e.g.}\ through the solution of the
Bethe-Goldstone equation.  The familiar idea underlying the infinite sum is
that the hard short-range core that makes the nucleon-nucleon interaction
intractable can be treated exactly, at least at the two-body level, by
nonperturbatively admixing into the wave function intermediate two-particle
states with arbitrarily high-energy.  The effective low-energy interaction
$G[V]$ that results has a soft core because the effects of short-range physics
have already been accounted for in the ladder sum.  The argument $V$ in the $G$
matrix is meant to reinforce the fact that $G$ depends on the ``bare
interaction'' $V$.

\begin{figure}[t]
\includegraphics{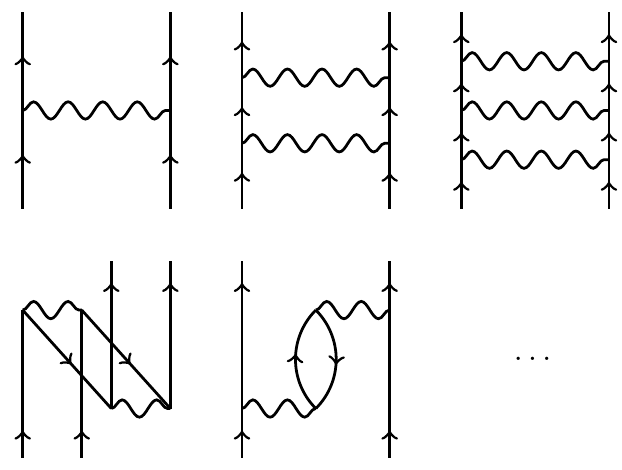}
\caption{\label{fig:fig2}Low-order diagrams (in $G$) for the effective interaction
$H_\mathrm{eff}$.}
\end{figure}

After this nonperturbative construction of the $G$ matrix, one can use
perturbation theory in $G$ to add the effects of states that are at low energy
but still outside the valence space, \emph{i.e.}\ in the intermediate space of
the double-partitioned set.  The diagrams in Fig.\ \ref{fig:fig2} include all
such effects up to second order in $G$ (with the exception of those produced by
tadpole and one-body graphs, which are commonly omitted), and some third-order
effects.  In this figure, the upward-going lines represent low-lying particle
states (including the valence levels, as long as they do not lead to
intermediate denominators with zero energy) in the gap between the Fermi
surface and the high-energy levels .  Downward going lines represent ``hole''
states that correspond to the vacating of levels below the shell-model space.

Typically, more complicated graphs, including the folded ones, are included
alongside the graphs in the figure.  Even then,  problems with convergence,
three- and higher-body operators that are generally too complicated to include,
etc., mean that the resulting interaction often must be modified
phenomenologically, especially in the monopole-monopole channel
\cite{duf96,cau05}.  Sometimes, however, the perturbation theory by itself is
enough to produce a pretty good interaction \cite{hjo95,cor08}.  The recent
shell-model calculations of double-beta decay in Ref.\ \cite{cau08} were based
on a tuned version of such an interaction.

A similar procedure can be followed to evaluate the matrix elements of any
two-body operator\footnote{Previous work has focused almost entirely on
one-body operators, however.} $\mathcal{M}$.  The expansion of equation
(\ref{eq:bhwf}) leads to a set of diagrams for the effective operator in Eq.\
(\ref{eq:meff}) in which a horizontal line representing the bare operator
appears once alongside an arbitrary number of interaction lines
\cite{bra67,ell77}.  The denominator of Eq.\ (\ref{eq:meff}) gives rise, in
addition, to norm and overlap diagrams that determine a special basis in which
the effective operator should be represented.  We have, however, evaluated the
most important of these diagrams (they are given by the derivative with respect
to the unperturbed energy of the effective-interaction diagrams) and found them
to contribute at most a few percent. We therefore won't include their
complicated but small effects here.

\begin{figure}[t]
\includegraphics{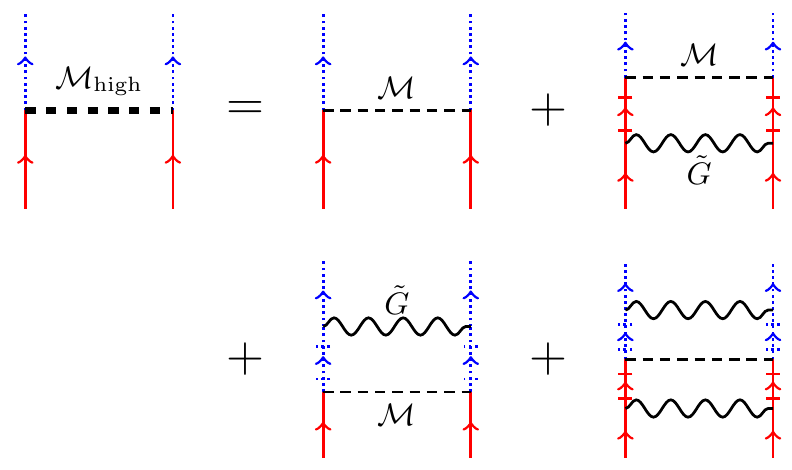}
\caption{\label{fig:fig3} (color online) An effective double-beta operator that
accounts for short-range correlations in the nuclear wave function.  Solid
lines (red online) are neutrons and dotted line (blue online) are protons. The
symbol $\tilde{G}$ represents the extension of $G$ to the $Q$ space.} 
\end{figure}

To obtain the effective decay operator, we begin by summing all diagrams with
two particles excited to high energies in which one horizontal line in each
diagram is the operator $\mathcal{M}$ rather than the interaction.  We denote
the result of this nonperturbative sum, which is completely analogous to the
$G$ matrix,  by $\mathcal{M}_\mathrm{high}$.  Sequences of 1, 2, \ldots
interaction lines either before or after the operator insertion can be
separately summed, \emph{i.e.}\ replaced by $G$ matrices (or, more, precisely,
the similar ladder sum $\tilde{G}$ for which the outgoing states are
high-lying).  If $\mathcal{M}$ is one of the double-beta operators appearing in
equations (\ref{eq:approxme}), then two neutron lines become proton lines
whenever it acts, and the ladder sum reduces to the four diagrams in Fig.\
\ref{fig:fig3}.  Thus, the solid lines (red online) in the figure represent
neutrons and the dotted lines (blue online) represent protons.  

Since these diagrams involve only $T=1$ states, the their sum (with small
Coulomb effects neglected) can be calculated simply from a $G$-matrix code
through the trick
\begin{equation}
\mathcal{M_\mathrm{high}}= \frac{d}{d\lambda} G[V_{T=1}+\lambda
\mathcal{M}']\big|_{\lambda=0} \,,
\label{eq:trick}
\end{equation}
where $\mathcal{M}'$ is the charge-conserving version of the charge-changing
operator in Eqs.\ (\ref{eq:f}) or (\ref{eq:gt}), obtained by removing the
isospin raising operators.  In other words, we can calculate matrix elements of
$\mathcal{M}_\mathrm{high}$ by computing the $G$ matrix corresponding to the
interaction $V+\lambda \mathcal{M}'$.  The derivative filters out all graphs
except for those that have a single double-beta line replacing an interaction
line.  The difference between the matrix elements of $\mathcal{M}$ and
$\mathcal{M}_\mathrm{high}$ give us a rigorous measure, at least for
two-valence-nucleon systems, of the effects of short-range correlations in
double-beta decay (up to the few percent due to norm diagrams).

\begin{figure}[b!]
\includegraphics{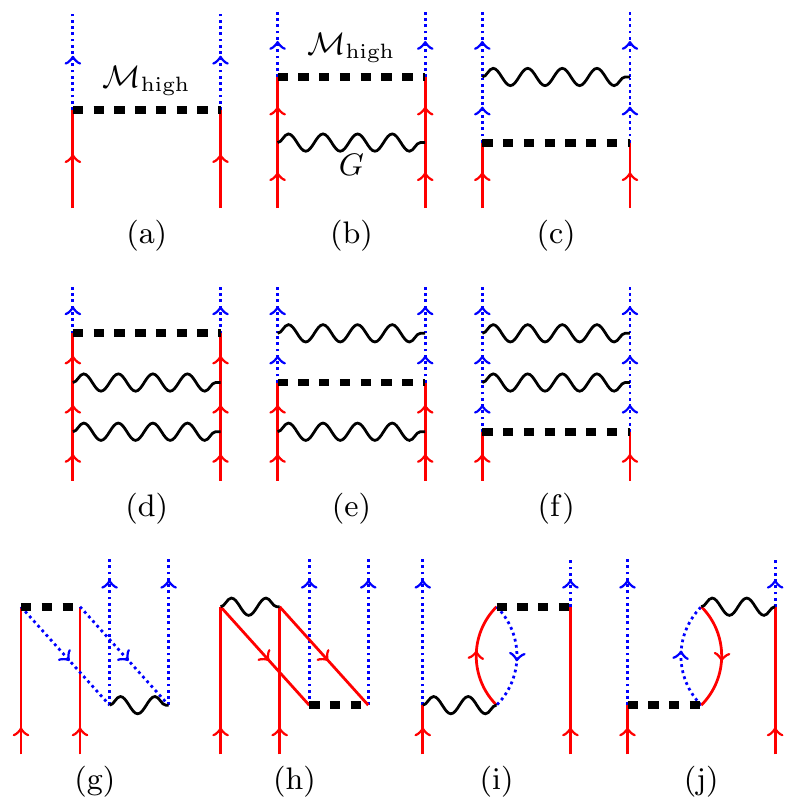}
\caption{\label{fig:fig4}(color online) Low order diagrams (in $G$)
contributing to the final effective double-beta operator.  The thick dashed
line is $\mathcal{M}_\mathrm{high}$, expressed diagrammatically in Fig.\
\ref{fig:fig2}.}
\end{figure}

Having constructed $\mathcal{M}_\mathrm{high}$ to include short-range two-body
correlations, one can use it together with the $G$ matrix to calculate the
additional renormalization from low-lying excitations.  We do so by replacing
one $G$-matrix line in each of the diagrams in Fig.\ \ref{fig:fig2} by
$\mathcal{M}_\mathrm{high}$.  The resulting diagrams for
$\mathcal{M}_\mathrm{eff}$, all first order in $G$ except for second-order
ladders, appear in Fig.\ \ref{fig:fig4}.  Short-range correlations are included
at every vertex through the use of $G$ and $\mathcal{M}_\mathrm{high}$ in place
of $V$ and $\mathcal{M}$.  Most of these diagrams can be calculated through the
trick in equation (\ref{eq:trick}).  Only the core-polarization graphs (the
last two in the figure) must be treated explicitly.  Those graphs are
essentially different from the corresponding effective-interaction graph in the
identities (neutron or proton) of the particles and holes involved.  One cannot
see the difference explicitly in our figures because Fig.\ \ref{fig:fig2} does
not distinguish between neutrons and protons.

\subsection{Using the operator in $^{82}$Se}

The shell-model calculations of Ref.\ \cite{cau08} used an interaction that was
constructed largely through the effective-interaction theory discussed above
(though it included more diagrams).  The Bonn-C nucleon-nucleon interaction
\cite{mac89} was the starting point for the ladder-diagrams that enter the $G$
matrix.  After summing all diagrams through third order in $G$ and the folded
diagrams based on those, the authors adjusted the interaction by fitting
certain components to spectra.

To be as consistent as possible with the calculations of Ref.\ \cite{cau08} in
constructing our effective decay operator for $^{82}$Se, we use the same Bonn-C
interaction, the same valence space ($f_{5/2}pg_{9/2}$), the same oscillator
parameter ($b=82^{1/6}$ MeV), and the same average energy ($\mybar{E}$ = 10.08
MeV) as that reference throughout.  After calculating the $G$ matrix and its
extension to the $Q$ space, we evaluate $\mathcal{M}_\mathrm{high}$ in Fig.\
\ref{fig:fig3} and the more involved effective-operator diagrams in Fig.\
\ref{fig:fig4}.  Our only modification to the standard procedure is to prohibit
intermediate particles in diagrams (i) and (j) from occupying levels that are
essentially full in $^{82}$Se and $^{82}$Kr; these Pauli-forbidden
contributions, if included, would be canceled by higher-order diagrams that we
do not evaluate here.  Finally, we combine the matrix elements of the effective
two-body operator (at several stages of approximation) with the 65 independent
two-body ground-state-to-ground-state transition densities from the shell-model
calculation of Ref.\ \cite{cau08} to obtain a transition matrix element for the
decay $^{82}$Se $\longrightarrow ^{82}$Kr.

\section{Results}

\subsection{High-energy states and short-range correlations}

Before looking at long-range corrections, we report the effects of the ladder
diagrams shown in Fig.\ \ref{fig:fig3}.  The treatment of short-range
correlations these diagrams represent is completely well defined; one knows
exactly what it includes and what it omits, and there is no double counting.
To look at the spatial structure of the correlations we define a two-body
double-beta correlation function $C^{GT}(r)$ by making the substitution
$H(r_{ab},\mybar{E}) \longrightarrow H(r_{ab},\mybar{E}) \,\delta(r-r_{ab})$ in
the Gamow-Teller transition operator.  

\begin{figure}[t]
\includegraphics[scale=.35]{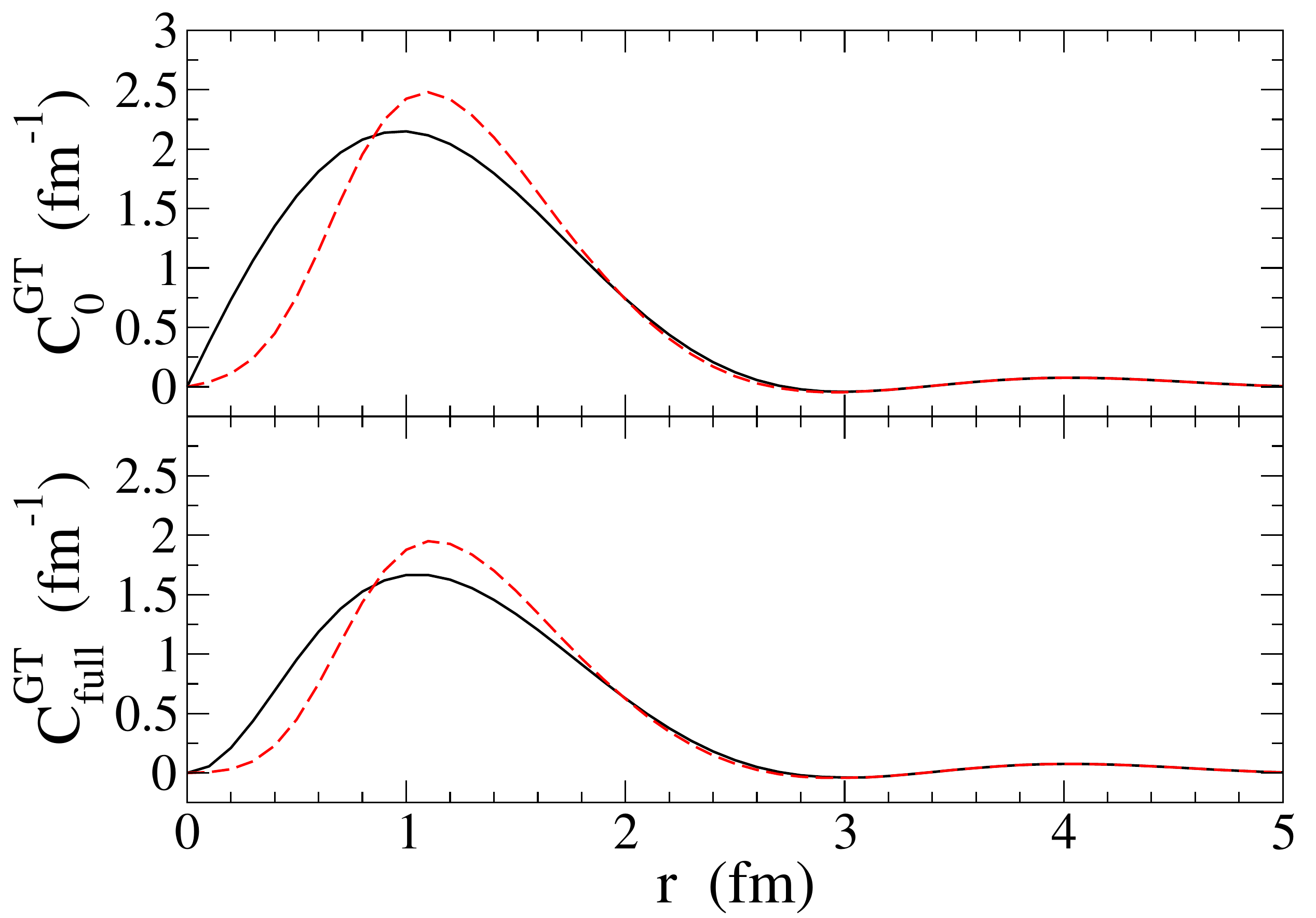}
\caption{\label{fig:fig5} (color online)  The radial distribution $C^{GT}(r)$
that, when integrated, produces the Gamow-Teller $0\nu$ matrix element for the
decay of $^{82}$Se.  The top panel corresponds to the simple operator given in
Eq.\ (\ref{eq:gt}) and the bottom panel includes momentum-transfer dependent
form factors and forbidden operators (see text). The solid lines correspond to
the bare operator and the dashed lines to the effective operator generated by
summing the high-energy ladders in Fig.\ \ref{fig:fig4}.}
\end{figure}

Figure \ref{fig:fig5} displays the results for the decay of $^{82}$Se with the
shell-model transition densities mentioned above, when the boundary between
``high-energy'' single-particle states (ruled in the diagrams) and lower-energy
states lies 4 $\hbar\omega$ above the valence $fp$ shell.  The results change
only very slowly as the boundary is moved up from that point.  The function in
the top panel, labeled with the subscript $0$, corresponds to the simple GT
operator of Eq.\ (\ref{eq:gt}); the function in the bottom panel includes
modifications to that operator from the weak nucleon form factors and
higher-order terms in the weak current (see, \emph{e.g.}, Refs.\ \cite{sim08}
and \cite{sim09} for definitions).    

\begin{figure}[b]
\includegraphics[scale=.35]{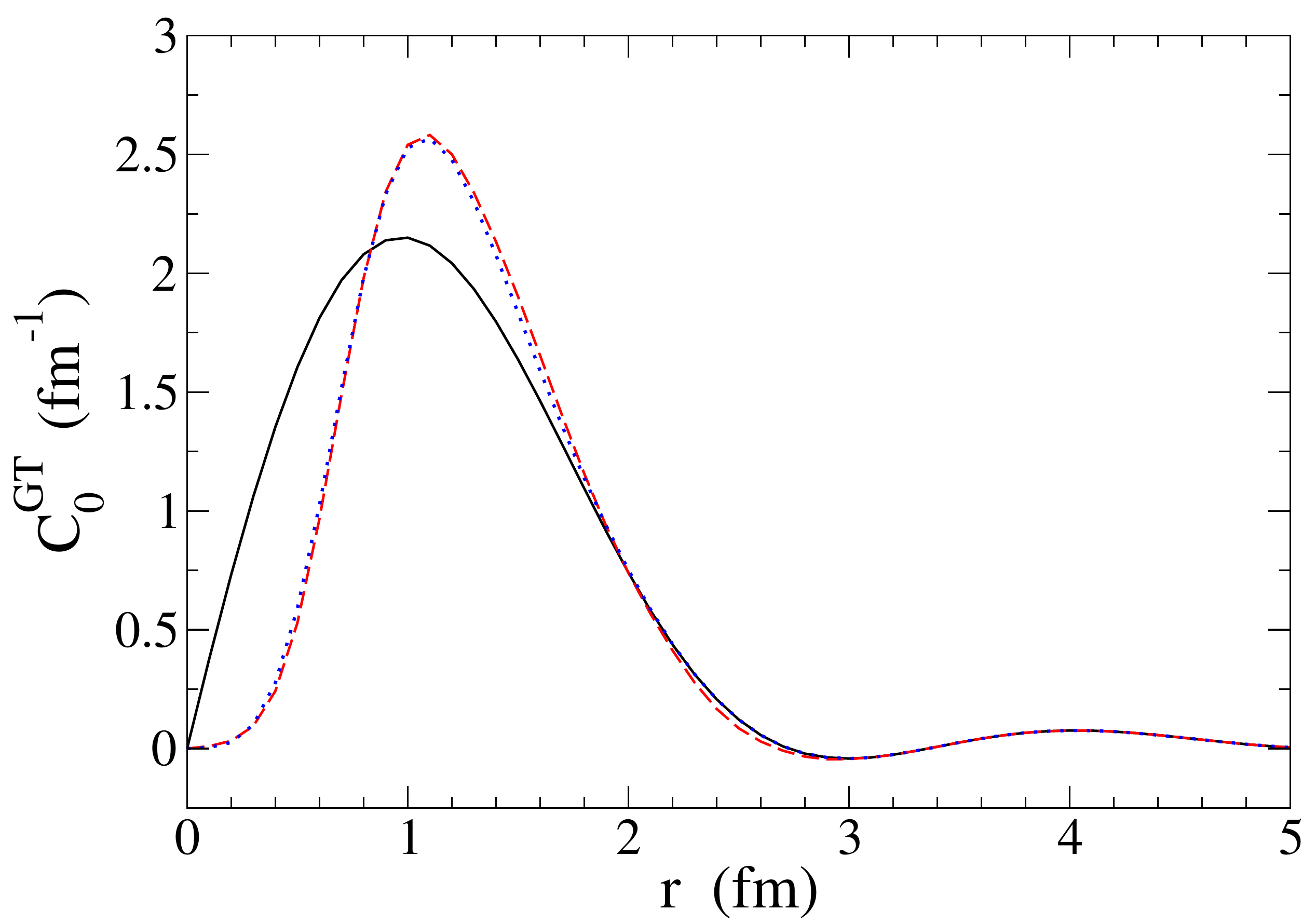} 
\caption{\label{fig:fig6} (color online) Same as the top panel of Fig.\
\ref{fig:fig5}, but with the dashed line from a calculation with the Argonne
V18 potential in place of Bonn-C.   The dotted line, nearly indistinguishable
from the dashed line, comes from a fit to coupled-cluster results \cite{sim09}
with the Argonne potential.}
\end{figure}

Both panels show the suppression of short-range contributions by the ladders,
though the suppression is weaker in the lower panel because the nucleon form
factors cut out some short-distance neutrino exchange by themselves.  But in
neither case is the matrix element reduced very much by the correlations:  for
the simple operator the reduction is about 8\%, and for the full operator it is
less than 3\%.   The reason, as the figure shows, is that probability density
is shifted from very short distances to around $r= 1$ fm, where the function
$H(r,\mybar{E})$ is still large. 

This behavior is similar to that found in Ref.\ \cite{sim09}, which constructs
correlated wave functions by solving coupled-cluster equations at the one and
two-body level.  There are some differences between that treatment of
short-range effects and ours:  the coupled-clusters method includes
Hartree-Fock-like effects, sums hole-hole ladders as well as particle-particle
ladders, and has no double partitioning.  Nevertheless, both methods include
much of the same physics and should yield similar results.  To test the
similarity we repeated our calculation with the Argonne V18 interaction used if
Ref.\ \cite{sim09} in place of the Bonn-C interaction.  Figure \ref{fig:fig6}
shows the results for the simple-operator GT correlation function discussed
above along with the corresponding coupled cluster results, which we generated
by using the phenomenological Jastrow-function fit reported in Ref.\
\cite{sim09}.  The two correlation functions are almost indistinguishable.
These results, alongside those of Ref.\ \cite{kor07,kor07a}, mean that the
phenomenological Jastrow function from Ref.\ \cite{mil76}, used reflexively for
a long time, almost certainly overestimates the quenching due to short-range
correlations. 

\subsection{Longer-range correlations}

When the correlations induced by the diagrams in Fig.\ \ref{fig:fig4} are added
to the short-range effects discussed above, the matrix element changes
further.  Results for the Gamow-Teller part of the matrix element (equation
(\ref{eq:gt}), without form factors or forbidden currents) appear in table
\ref{tab:1}.

The table successively adds the results from three classes of diagrams:  the
high-energy ladders discussed above (labeled (a) in Fig.\ \ref{fig:fig4}), the
diagrams with the high-energy ladders embedded in lower energy ladders (labeled
(b) -- (f)), the 4-particle 2-hole diagrams (g) and (h), and finally the
core-polarization diagrams (i) and (j).  Each row corresponds to a different
boundary (measured in $\hbar \omega$ from the $fp$ shell) between the
``high-energy'' particle levels, denoted by ruled lines in the figures, and the
lower-energy levels.  All the diagrams except the last two in Fig.\
\ref{fig:fig4} are insensitive to this boundary if it is above about 4
$\hbar\omega$.  The core polarization graphs, by contrast still have not
converged at 8 (or even 9) $\hbar\omega$.  We are unable to carry the
calculation beyond that point.  Core-polarization graphs in the effective
interaction are notorious for converging very slowly, sometimes taking 20 or
more $\hbar\omega$ \cite{var73}.

\begin{table}[tb]
\caption{\label{tab:1}Renormalization of $M^{0\nu}_{GT}$ (from equation
(\ref{eq:gt})) for the decay \protect{$^{82}$Se $\longrightarrow ^{82}$Kr}.}
\begin{tabular*}{0.4\textwidth}{@{\extracolsep{\fill}}c|@{\hspace{.5cm}}ccccc}
 boundary& bare& (a) & (a)-(f)  & (a)-(h)
  & all\\
 \hline
   $4 \hbar\omega$ &  $3.33$ & $3.07$ & $4.15$ & $5.38$
 & $3.05$\\
  $5 \hbar\omega$ & $3.33$ & $3.06$ & $4.17$ & $5.39$
 & $3.15$ \\
  $6 \hbar\omega$ & $3.33$ & $3.05$ & $4.16$ & $5.39$
 & $3.21$\\
  $7 \hbar\omega$ & $3.33$ & $3.06$ & $4.17$ & $5.39$
 & $3.28$\\
  $8 \hbar\omega$ & $3.33$ & $3.06$ & $4.17$ & $5.39$
 & $3.35$
\end{tabular*}
\end{table}

The table shows several things.  First, the short-range correlations, as
discussed previously, damp the bare matrix element by about 8\%.  Second, the
ladder and 4-particle-2-hole contributions then increase the matrix element by
about 75\%. These graphs contain pairing matrix elements that promote particles
into unoccupied levels, unblocking the transition.  Finally, the
core-polarization diagrams decrease the matrix element so that after summing
particle-hole configurations up to 8 $\hbar\omega$ our matrix element is only
marginally bigger than the bare version.  The core-polarization graphs contain
the neutron-proton interaction, the correlations from which are known to
counteract the effects of pairing.  Cutting off the sum at 8 $\hbar \omega$,
however, probably exaggerates the size of this counteraction; the bulk of the
effect comes from low energy levels, and contributions from higher-energy
levels actually increase the matrix element again.  Thus, our full result (in
the column labeled ``all'') grows with the boundary between low and high-energy
states, and is still growing at our maximum value.

All these statements remain true when true when we include the Fermi term in
Eq.\ (\ref{eq:f}) --- resulting in a total $M_{0\nu}$ of 3.95 at 8
$\hbar\omega$ vs.\ the bare value 3.78 --- or add the effects of form factors.
We have not included forbidden currents in the full calculations, but do not
expect them to change the pattern reported in the table. 

As just noted, the effects of the pairing and neutron-proton correlations
cancel each other to a significant extent.  The cancellation resembles what
happens in the QRPA, which includes a portion of the effects calculated here
through the use of a relatively large single-particle space.  When all is said
and done, our result, is a bit larger (and continuing to  grow at $8
\hbar\omega$) than result at the bottom of column (a).  That number is the bare
matrix element corrected for short-range correlations, like the results
reported in Refs.\ \cite{men08} and \cite{cau08}.   The use of our effective
operator thus improves the agreement between the shell model and QRPA, though
the convergence issues keep us from saying by exactly how much.  One would
expect something similar for $^{76}$Ge, which has only six fewer nucleons and
is treated in the same model space.  

Of course, the large corrections from individual graphs mean that higher order
corrections may be sizeable as well.  One cannot, therefore, take the results
of this first-order estimate too seriously.   In particular, it would not be
correct to conclude that the near cancellation between pairing and
neutron-proton correlations persists to higher order and/or when many-body
diagrams are included.  The large individual components indicate that the
contributions of states outside the model space are significant and may or may
not cancel one another in a better calculation.  A more reliable
nonperturbative evaluation of the corrections is very important, and we are
working in that direction.

\begin{acknowledgments}

We acknowledge useful discussions with Morten Hjorth-Jensen and thank Alfredo
Poves for supplying shell-model densities.  This work was supported in part by
the U.S.\ Department of Energy under Contract No.\ DE-FG02-97ER41019 with UNC
and Contract No.\ DE-AC05-00OR22725 with UT-Battelle, LLC.
\end{acknowledgments}

\end{document}